\documentclass[preprintnumbers,floats,prd, twocolumn, longbibliography,floatfix, nofootinbib]{revtex4-1}
\usepackage{graphicx}
\usepackage{float}
\usepackage{subfigure}
\usepackage{dcolumn}
\usepackage{bm}
\usepackage{amsmath}
\usepackage{amsthm}
\usepackage{multirow}
\usepackage{enumerate}
\usepackage{amsfonts}
\usepackage{ifthen}
\usepackage{psfrag}
\usepackage{slashed}
\usepackage{hyperref}
\usepackage{gensymb}
\usepackage[utf8]{inputenc}
\usepackage{color}
\usepackage{subfigure}
\usepackage{ulem}
\usepackage[utf8]{inputenc}

\newcommand{\vev}[1]{\left\langle #1 \right\rangle}

\newcommand{\be}{\begin{equation}}
\newcommand{\ee}{\end{equation}}
\newcommand{\bea}{\begin{eqnarray}}
\newcommand{\eea}{\end{eqnarray}}

\newcommand{\mos}{{M\"ossbauer}~}

\begin{document}

\title{A M\"ossbauer Scheme to Probe Gravitational Waves}

\author{Yu Gao}
\email{gaoyu@ihep.ac.cn}
\affiliation{Institute of High Energy Physics, Chinese Academy of Sciences, Beijing, 100049, China}

\author{Huaqiao Zhang}
\email{zhanghq@ihep.ac.cn}
\affiliation{Institute of High Energy Physics, Chinese Academy of Sciences, Beijing, 100049, China}

\author{Wei Xu}
\email{xuw@ihep.ac.cn}
\affiliation{Institute of High Energy Physics, Chinese Academy of Sciences, Beijing, 100049, China}

\begin{abstract}
Under the local gravitational field, perturbations from high-frequency gravitational waves can cause a vertical shift of the M\"ossbauer resonance height. Considering a stationary scheme with the $^{109}$Ag isotope, we demonstrate that the extremely high precision of M\"ossbauer resonance allows for competitive gravitational wave sensitivity from KHz up to above MHz frequencies. M\"ossbauer resonance can offer a novel and small-sized alternative in the quest of multi-band gravitational wave searches. The presence of the static gravitational field plays essential role in the detection mechanism, isotope selection and sensitivity forecast. The proposed stationary scheme's sensitivity has the potential of significant improvement in a low-gravity environment.
\end{abstract}

\maketitle


\section{Introduction}
\label{sect:intro}

Shortly after its first discovery in 1958~\cite{Mossbauer:1958wsu,Mossbauer2}, the \mos  resonance played an important role in the early quest of testing relativity~\cite{PhysRevLett.3.439,PhysRevLett.3.439,PhysRevLett.3.556,Barit} due to its ultra-high frequency precision. A series of laboratory measurements were successfully carried out at the Atomic Energy Research Establishment~\cite{PhysRevLett.4.165,Cranshaw_1964}, at the tower experiment in Jefferson Physical Laboratory~\cite{Pound:1960zz,Pound:1964zz}, and famously demonstrated a height-induced $2ghc^{-2}\sim 4.905\times 10^{-15}$ frequency shift in 1965~\cite{Pound:1965zz}, confirming Einstein's equivalence principle. Early \mos test for the equivalence principle also include the measurements in non-inertial systems~\cite{Kundig1963,Champeney1961}. Later \mos experiments are carried out with higher precision, for instance the angular measurement with $^{67}$Zn~\cite{KATILA198151} and inside a cryostat~\cite{Potzel1992}, null-redshift tests with a differential \mos scheme~\cite{Vucetich:1988uw,DeFrancia:1992em} and for displacement sensing~\cite{Ikonen1991}, etc. 
For comprehensive history reviews, see Ref.~\cite{Gonser1981,Hentschel_review} and references therein. 
Over the decades, tests of general relativity gradually shifted toward other advanced techniques: most importantly the high-precision timing with clocks~\cite{PhysRev.104.11,Hafele1972,Alley1537236,Sappl1990}, maser experiments~\cite{osti_6981725,PhysRevD.27.1705}, gyroscopes~\cite{Jeckins1969AJ} such as the recent Gravity Probe B~\cite{Everitt:2011hp}, long-distance Michelson interferometry with LIGO~\cite{Abramovici:1992ah} and VIRGO~\cite{Mours:1993rn}, as well as future space programs such as LISA~\cite{Danzmann:1997hm}, TianQin~\cite{TianQin:2015yph} and Taiji~\cite{Hu:2017mde}. Various reasons might have caused fewer \mos applications in major gravity-test programs~\cite{Hentschel_review}. Nevertheless, the extreme precision in resonance frequency keeps inciting novel ideas, such as \mos superradiance with Rhodium~\cite{cheng2007rhodium}, resonance with low-frequency nuclear spin flips~\cite{Hannon1989XrayRE,Cheng:2007wb} and \mos rotor experiment~\cite{Corda:2019agu}, etc.

\begin{figure*}[t]
    \centering
    \includegraphics[scale=0.5]{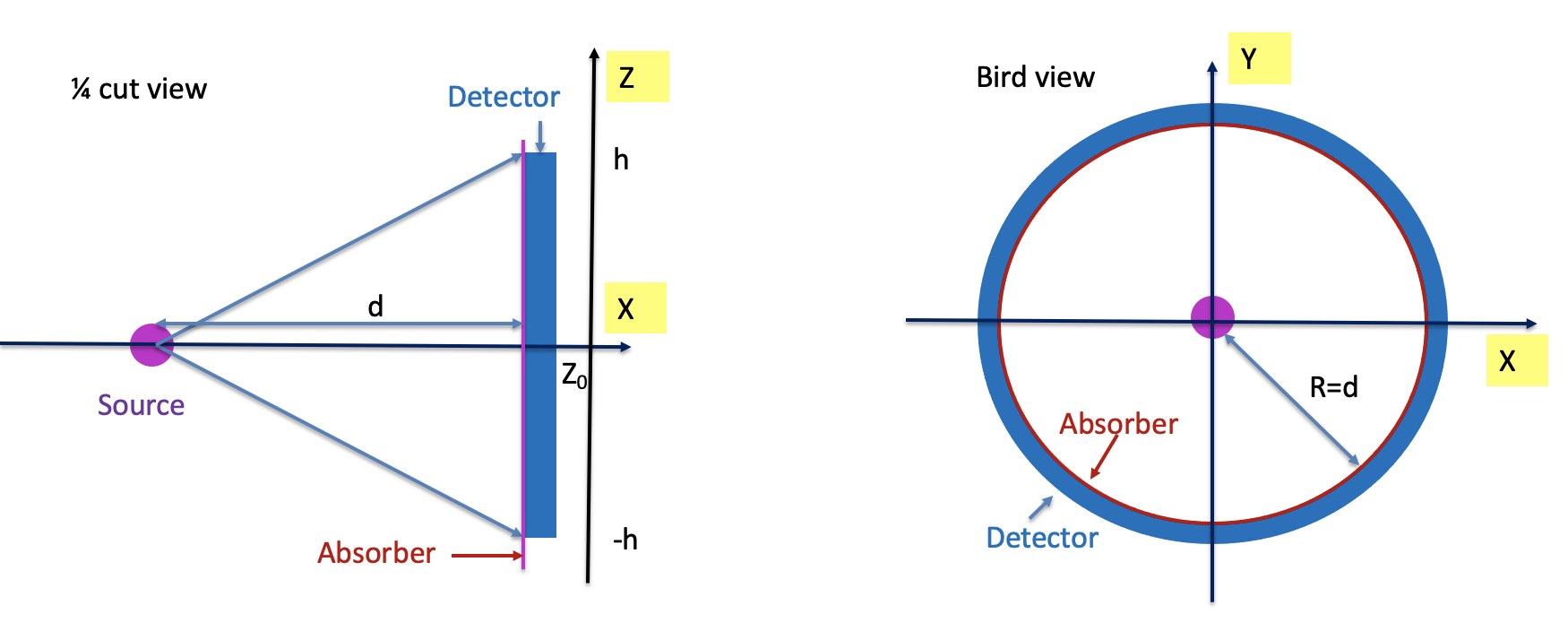}
    \caption{Stationary measurement: a frequency shift $\delta f$ causes the resonance point to move vertically at the detectors located on a horizontal circle. Detector dimensions are exaggerated for illustration.}
    \label{fig:scheme}
\end{figure*}

Since the LIGO discovery~\cite{LIGOScientific:2016aoc} in 2016, and recently indicated by nano-Hz pulsar timing observations~\cite{NANOGrav:2020bcs,NANOGrav:2023gor,Xu:2023wog,Antoniadis:2023utw,Reardon:2023gzh}, the search for gravitational waves (GW) has become a much heated frontier, where many new search proposals have emerged, see Refs.~\cite{Ballmer:2022uxx,Carlesso:2022pqr} for recent reviews. If GWs exist as a background, photon propagation will experience frequency fluctuations. Interest in a direct \mos detection of such a GW effect has undoubtedly long existed~\cite{kaufmann1970}, as the \mos resonance is in principle sensitive to a photon energy fluctuation at the order of $10^{-15}$ or even smaller.
Under laboratory conditions, as the emitter and the absorber are spatially close, a strain difference $\Delta h$ between the space-time points the photon's emission and absorption is then required to create a frequency shift.
This typically requires high-frequency GWs with a wavelength shorter than the \mos baseline length $d$, or $f_{\rm GW}> 4c/d\sim (1.2{\rm m}/d)$ GHz. In principle lower frequencies may also contribute a non-zero $\Delta h$, with a linear fraction of $\lambda_{\rm GW}/4d$ suppression on the GW amplitude. Besides a stochastic GW background, there are no well-known high-frequency sources in the standard particle physics and cosmology models. Nevertheless, new physics such as primordial black hole mergers~\cite{Kocsis:2017yty,Raidal:2018bbj,Gow:2019pok}, superradiance around Kerr black holes ~\cite{Ternov:1978gq,Zouros:1979iw,Brito:2015oca}, light bosonic dark matter~\cite{Hui:2021tkt,Antypas:2022asj}, plus other exotics, predict potential coherent GW sources in the MHz-GHz range, and a number of novel detection methods have been proposed~\cite{Aggarwal:2020olq}. Such high-frequency GW can serve a potential candidate for \mos observation.

Several isotopes~\cite{MEDC} are known for their particularly sharp \mos lines. To name a few, the 14.4 keV transition of $^{57}$Fe has a width  $\delta\lambda/\lambda\sim 7\times 10^{-13}$; the 93 keV transitions of $^{67}$Zn has  $\delta\lambda/\lambda\sim10^{-15}$; $^{73}$Ge at $3\times 10^{-14}$; the 88 keV transition width of $^{109}$Ag in principle can be as low as $10^{-22}$, and several other isotopes such as $^{103}$Rh, $^{107}$Ag and $^{189}$Os also have extremely narrow natural linewidth~\cite{Davydov2001}. In practice, the achievable linewidth must account for various line shifts from second-order Doppler effect~\cite{PhysRevLett.4.274}, inhomogeneities in material's chemical composition~\cite{PhysRevLett.7.405} and mechanical vibrations~\cite{Potzel1992}, etc. $^{57}$Fe is the most commonly used isotope to date. Mature techniques allow an improved frequency sensitivity at a fraction of the natural width.
For $^{109}$Ag, a sensitivity at 30 times of its natural width was realized in an attempt in 1979~\cite{Wildner1979ANA}, and the resolution has been advanced over a series of experiments~\cite{10.1117/12.943893,Hoy:1990,Alpatov:1996bd,Alpatov:2007LF}, with the inclusion of gravity-induced effects~\cite{Bayukov:2009Ag}. Given these considerations on the frequency resolution, we will discuss a conceptual \mos experiment, and access the prospects of measuring the recoil-less photon's frequency shift arising from passing-by gravitational waves.

\bigskip

\section{A Stationary Scheme}
\label{sect:scheme}

\begin{figure}[t]
    \centering
    \includegraphics[scale=0.39]{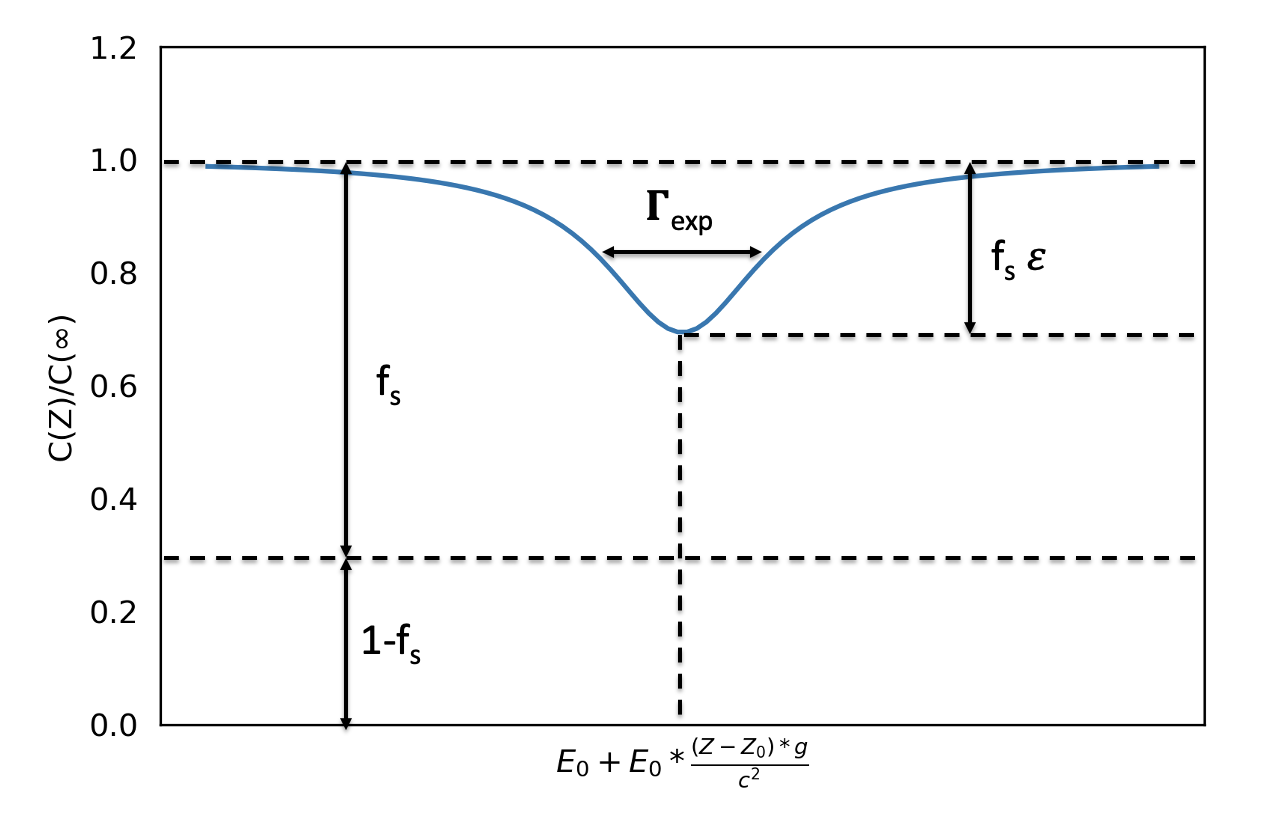}
    \caption{Expected counting spectrum normalized to far away of recoil-free absorption peak.The $x$-axis represents the gravitational energy shift by height difference $Z-Z_0$. The $\Gamma_{\rm exp.}$ is typical twice of the natural width for thin layers ($t\ll 1$), yet it increase significantly with thick absorbers.}
    \label{fig:spectrum}
\end{figure}

As gravitational waves have their own frequencies, we consider a static measurement scheme\footnote{stationary measurement is also known to alleviate vibration uncertainties~\cite{Vucetich:1988uw,Potzel1992}} that uses the difference in the vertical distance among detectors to resolve any frequency shift of the $\gamma$ rays emitted from the source. Let us assume the source is stationary at a vertical position $Z_S$, and the source's emission line is narrow and unsplit, with a natural width $\Gamma$. The measured lineshape is Lorentzian with a central resonance energy of $E_0$. Denoting the total photon emission rate of the source as $\dot{N}_0$, then the differential number of recoil-free (RF) photons with energy $E_{\gamma}$ per unit energy and time is
\be 
\frac{{\rm d}N_{\rm RF}(E)}{{\rm d}E ~{\rm d} t} = \dot{N}_0 f_{s}\cdot  \frac {\Gamma /2\pi} {[E-E_0]^2 + (\Gamma /2)^2 }~,~~
\label{eq:p_emmit}
\ee
where $f_S$ is the recoil-less fraction of the source emission. Now let's consider an absorber with a mean energy of $E_0$ between the excited and ground states, same as that in the source, is placed between the source and a photon detector. Here the detector plays the role of a photon counter that observes the variation of the photon flux through the absorber. A number of stationary absorbers and detectors are fixed along a horizontal ring with same distance $d$ to the sources, and the horizontal ring is close to the height of the source, see Fig.~\ref{fig:scheme} for illustration. We require the detector to have good spatial resolution and the finite size of the absorbers and detectors will allow them to cover a small vertical range. In this static configuration, the \mos transmission integral can be derived by replacing the  Doppler-shift term with a height-induced correction:
\bea
\label{formula:spectrum}
C(Z)  &=& \dot{N}_0 ~{e}^{-\mu_e t'} \cdot \left[(1-f_S) + \vphantom{\frac{XX}{XX}}\right. \label{eq:a_emmit}\\
&&\left.\int_{-\infty}^{\infty} f_S\xi(Z_S,E_0)\cdot {\rm e}^{-t \xi(Z,E_0+\Delta E_0)\Gamma /2\pi } {\rm d}E \right], \nonumber \\
{\rm}~\xi(Z,E_0)&\equiv& \frac{\Gamma /2\pi}{[E-g(Z-Z_S)E-E_0]^2 + (\Gamma /2)^2}, \nonumber
\eea
and $\Delta E_0$ denote the intrinsic shift between the source and the absorber~\cite{Mossbook-GET2011} that is compensated for via a small height difference within the covered range of detectors. Namely, at the resonance point we would expect $g\vev{Z-Z_S}E=-\Delta E_0$ and we denote this central height as $Z_0\equiv \vev{Z}$ in the rest of the paper. We also adopt the natural unit system where $\hbar=c=1$. The quantity ${e}^{-\mu_{e} t'}$ is the mass attenuation factor, in which $t'$ denotes the absorber thickness a.k.a. the area density, 
and $\mu_e$ is the total mass absorption coefficient of the absorber at resonant emission energy. $t=f_AN_M\sigma_0$ is the effective resonant absorption depth, where $f_A$ is the fraction of recoil-free absorption at the absorber, $N_M$ is the absorber's number of \mos nuclei per unit surface area that increases with the absorber thickness, and $\sigma_0$ is the maximum total resonant cross section. Same as in the conventional \mos case~\cite{Margulies1961TRANSMISSIONAL}, the height-corrected $C(Z)$ can also be expanded into a more convenient parametrization:
\be 
C(Z)=\dot{N}_0 ~{e}^{-\mu_e t'} \left\{1-f_S~\epsilon\cdot\frac{ \Gamma_{\rm exp}^2}{[g(Z-Z_0)E_0]^2+\Gamma_{\rm exp}^2}\right\},
\label{eq:parametrization}
\ee
which is plotted in Fig.~\ref{fig:spectrum}. Here $\epsilon$ is resonant absorption fraction of absorbers, $\Gamma_{\rm exp}$ is the observed width behind the absorber, and these parameters nontrivially depend on the absorber thickness $t$. In Fig.~\ref{fig:spectrum}, the section labeled by $f_S$ represents the recoil-free emission fraction, the  $f_S\epsilon$ section represents the total recoil-free absorption fraction and the $1-f_S$ section is the recoiled emission fraction. 

Generally one would need to balance between a larger absorption fraction with the mass attenuation. Dedicated studies~\cite{long1983ideal} showed that mass attenuation near $\mu_e t'\approx 2$ will give the best sensitivity. For natural silver composed of $^{107}$Ag (52\%) and $^{109}$Ag (48\%), this corresponds to a thickness of 0.93 mm silver for the 88 keV line and the effective resonant absorption depth is around 7 for $^{109}$Ag. In the following, we consider a benchmark case with $t=8$ for higher concentration of $^{109}$Ag in the absorber, and correspondingly $\epsilon \approx~0.8$ and $\Gamma_{\rm exp.} \approx~4.1~\Gamma$~\cite{RANCOURT199185,Mossbauer1960,MARGULIES1961131}. Since the 88 keV line of $^{109}$Ag is very narrow, its resonance height range $g^{-1} \Gamma_{\rm exp}/E_0$ falls in a reasonable detector size, and the spatial location of maximal absorption can be resolved when $\Gamma_{\rm exp.}$ is narrower than the gap between adjacent peaks. In case $\Delta E_0$ varies between different absorbers, the exact central location $Z_0$ can be calibrated by carefully adjusting the height of each detector. Here we will not go depth with multi-peaked spectral analysis and assume one resolvable resonance peak for this proof-of-principle study. We will perform simulations on the detector sensitivity and discuss more detailed requirements later in Section~\ref{sect:signal}. In addition, the experimentally resolved $\Gamma_{\rm exp.}$ may suffer a broadening factor~\cite{Bayukov:2009Ag}. We will show later that the spatial sensitivity scales only by the square-root of such a factor, and use $4.1~\Gamma$ as the benchmark.

\medskip

Now let us consider an additional frequency shift $\Delta f(t)$ entering the system, so that the vertical location of the resonance band will move accordingly, 
\be 
Z_0 \rightarrow Z_0(t) = Z_0 + g^{-1} \frac{\Delta f(t)}{f_\gamma},
\label{eq:h_movement}
\ee
and the movement of the resonance band $\Delta Z_0(t)\equiv Z_0(t)-Z_0$
can be measured by the observing the \mos absorption efficiency at the detector array. If spatial resolution is good, 
$\Delta Z_0(t)$ and correspondingly $\Delta f(t)$ can be measured to a sensitivity level better than the effective \mos linewidth $\Gamma_{\rm exp.}/E_0$. 

Here we would like to emphasize that this setup does {\it not} aim to compare the absolute height $Z_0$ at the maximal resonance, because the conditions at the source and the absorber can not be perfectly identical; the calibrated $Z_0$ value does not need to be the same for all detectors in different horizontal directions. Instead, we are interested in a (time-dependent) {\it variation in $Z_0$} as the signal for any additional frequency shift in our stationary system, such as those from gravitational waves.

\section{Gravitational wave signal}
\label{sect:GW}

To find out the response of our setup to GWs, consider a gravitational plane-wave along the $\hat{z}$ direction ($\vec{k}_{\rm GW}//\hat{z}$),
\be 
h = h_0~ \cos{(\omega t- \omega z)},
\ee
where we use the lower-case coordinates for the frame in which GW propagates along $\hat z$, not to be confused with the capital coordinates for the lab-frame where detectors are place on the horizontal $\hat X-\hat Y$ plane and resonance height shifts vertically along $\hat Z$. $h_0$ denotes for the magnitude of the GW strain, and it satisfies 
\be 
{\rm d}s^2 = {\rm d}t^2 - (1+h){\rm d}x^2 - (1-h){\rm d}y^2 -{\rm d}z^2.
\ee
As a photon propagates in the GW background, the photon will experience a difference in strain $h(t,\vec{x})$ at different space-time locations $(t,\vec{x})$, which causes a frequency shift on the order of $h$. The analytic expressions of the frequency shift between the source and detector has been derived in a number of early works~\cite{kaufmann1970,Grishchuk:1974jy,Estabrook1975,Hellings:1978vp,DeFelice:1979uu}, and also see~\cite{Faraoni:1991bk,Harte:2015ila,Noskov:2017} for more exotic circumstances. Here, we adopt the treatment in Ref.~\cite{Estabrook1975,Hellings:1981xc}, that the photon's frequency shift after its one-way propagation over distance $d$ in the direction of $(\theta, \phi)$ is given by
\be 
\frac{\Delta f_{}}{f_\gamma} =\frac{\ell^\mu \ell^\nu}{1-\cos\theta}[ h_{\mu\nu}^{\rm D} -h_{\mu\nu}^{\rm E}]
\ee 
where the superscripts $^{\rm D}$ and $^{\rm E}$ denote the 4-positions $(t,\vec{d})$ and $(t-d,\vec{0})$ at the detection and the emission of the photon. $t$ is the time at that photon reaches the detector, and we let it absorb the initial phase of the GW. $\ell^\mu=f_{\gamma}(1, \sin\theta\cos\phi,\sin\theta\sin\phi,\cos\theta)$ is the unperturbed propagation vector of the photon~\cite{Hellings:1981xc}. Here we will ignore the small $\sim{\cal O}(h)$ fluctuation in the photon's direction. Folding $\ell^\mu$ and $h_{\mu\nu}$ into the formula above, it can be rewritten into
\bea  
\frac{\Delta f_{}}{f_\gamma} &=& 2h_0 \cos^2\frac{\theta}{2}\cos{2\phi}\sin \left(\omega d \sin^2\frac{\theta}{2} \right) \label{eq:freq_shift} \\
&\cdot&\sin \left(\omega t -\omega d \cos^2\frac{\theta}{2} \right), \nonumber
\eea
where $\omega$ is the angular frequency of the GW. This frequency shift vanishes when the photon propagates exactly (anti)parallel to the GW's propagation direction. At low GW frequencies, or $\omega d \ll 1$, $\Delta f$ maximizes at $\theta\rightarrow \pi/2$, namely in the perpendicular direction of the GW.

\begin{figure}
    \centering
    \includegraphics[scale=0.75]{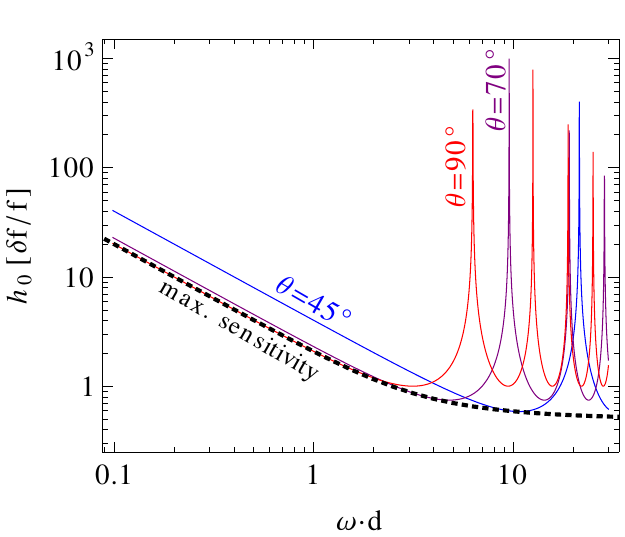}
    \caption{A single detector's GW strain sensitivity at distance $d$ in terms of effective \mos sensitivity on photon's energy shift. The $\theta$-labeled curves denote for detectors placed at $\theta=45^{\degree}, 70^{\degree}$ and $90^{\degree}$. The black dotted curve shows the maximal sensitivity floor by optimizing the angle $\theta$. Larger angular coverage with detectors will help approach to this limit. In this plot we choose $\phi=0$.}
    \label{fig:frequency_mod}
\end{figure}

At higher GW frequencies, $\omega d \gg 1$, however, this relation becomes more complicated: the amplitude in Eq.~\ref{eq:freq_shift} develops a series of `blind spots' at
\be 
\omega d \sin^2{\frac{\theta}{2}} = n\pi  , ~~ n=1,2,3...
\ee
where $\Delta f$ also vanishes. This means for a detector at a fixed direction and distance, its sensitivity is frequency-modulated in the high-frequency range, as illustrated by the peaks in Fig.~\ref{fig:frequency_mod}. High frequency GW with $\omega\gg d^{-1}$ will finds several insensitive angles between $0<\theta<\pi$. As a way out, multiple detectors at different directions can compensate for each other's blind frequencies. With our circular placement of detectors in Fig.~\ref{fig:scheme}, the incident GW at angle $\theta$ to the detector plane can be probed in the angular range $\theta\in (\theta,\pi-\theta)$ along the circle. The maximal frequency-shift, by optimizing the angles, is 
\be
\left.\frac{\Delta f_{}}{f_\gamma}\right|_{\rm max.} =\left\{
\begin{array}{cc}
    \frac{\omega d}{2} h_0, &\ \  ~ \omega d\ll1~\&~\theta\rightarrow \frac{\pi}{2}, \\
    \eta(\omega d)\cdot h_0, &\ \  ~ {~ \omega d>1, ~\rm 1^{st} ~max.}
\end{array}
\right.\label{eq:max_sen}
\ee
where $\eta(\omega d)$ is a frequency-dependent coefficient between 0.5 and 2, and it saturates to $\eta\approx 2$ in the high frequency limit. This optimal sensitivity is illustrated by the bottom curve (black-dotted) in Fig.~\ref{fig:frequency_mod}, and it is reached at the first maximum for $\theta >0$. The $h_0$ sensitivity is obtained by comparing to this maximal frequency shift within the observational angular then at a given incident GW direction. Note that for a given $h_{\mu\nu}$ pattern, both $\theta$ and $\phi$ vary along the circle. Therefore in principle our circularly-placed detectors can probe {\it both} the GW strain amplitude $h_0$ and the GW polarization angle $\phi$. An advantage with the circle-shape of detector placement is that the $\theta=\pi/2$ direction is always observed. When the incident GW is perpendicular to the (horizontal) circle's plane, there will be four maxima around the circle due to the $\cos2\phi$ dependence.

There is an important difference between the signal from a background GW and that from a static gravity field. In the GW case, the energy-shift is time-dependent, as clearly seen in Eq.~\ref{eq:freq_shift}, thus the signal must be resolved at a frequency no less than the GW frequency. This will lead to practical limitation on how frequently the detector's resonance status can be read out, which we will discuss next. Also, as increasing the GW frequency only gives an ${\cal O}(1)$ improvement on the maximal frequency shift, thus it is cost-effective for this experimental setup to target at the $\omega d \le {\cal O}$(10) regime.

\section{Detector requirements}
\label{sect:signal}

For signal detection, the sensitivity to is proportional to the maximal $Z$-shift of the absorption peak position caused by gravitational waves within a measurable time period and a reasonable spatial detection region. We will discuss a detector setup and estimate the stationary scheme's benchmark sensitivities based on governing factors such as typical detector specifics, \mos resonance fraction and source intensity, etc.

\medskip

(1) {\it Spatial resolution} and the resonance region size.
The natural width of the 88 keV emission line of $^{109}$Ag  is $2.3 \times 10^{-17}$ eV~\cite{MEDC}, and the perfect resonance width is magnified by a factor 2 due to both emission and absorption. In practice, the experimentally resolved width is broadened by smearing effects in the sample's material, including the effective absorber thickness~\cite{MARGULIES1961131,Cranshaw_1974,shenoy1974}. The broadening factor is 4.1 for an effective absorption depth $t= 8$. With such a configuration, the experimental 88 keV linewidth is expected to be $\Gamma_{\rm exp.}=1.9\times 10^{-16}$ eV, and it is equivalent to a vertical shift of $\delta Z = 20 \mu$m for an environmental $g =9.8 ~{\rm m/s^2}$ on the Earth's surface. The detector's spatial resolution is chosen to be half of the absorption peak's experimental size, namely 10 $\mu$m, and we would assume a good detection efficiency close to 100\% for the energetic X-ray photon. This small pixel should be possible via  R\&D with high-z detectors, such as with Cadmium telluride (CdTe) or Cadmium-Zinc telluride (CdZnTe)~\cite{LIMOUSIN200324}. An interesting possibility is one may find ways to reduce $g$, for instance, under space-borne environments. A smaller $g$ significantly increases the resonance $\Delta Z$ size. Taking $g=10^{-2}{\rm ~m/s^2}$ as an example, $\Delta Z$ for will be raised above the centimeter scale, and conventional X-ray detectors like NaI(Tl)/CsI(Na) phoswich detectors~\cite{liu2029high} can be capable of the task.

\begin{figure}[t]
    \centering
    \includegraphics[scale=0.55]{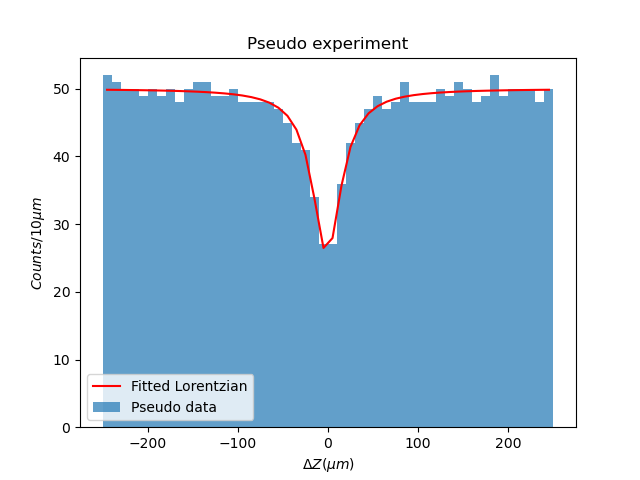}
    \caption{Simulated pseudo-experiment that measures the absorption Lorentzian peak position of experimental width of 20 $\mu m$, the bin width of X-axis is 10 ${\rm \mu m}$, 50 emitted 88 keV gamma rays in each bin, with recoil-free fraction of $f_S=0.6$, and absorption fraction of $\epsilon=0.8$. This pseudo-experiment gives the peak position accuracy of 0.67 ${\rm \mu m}$ that corresponds to a frequency shift sensitivity at $\delta{f}/{f} = 7.3 \times 10^{-23}$.}
    \label{fig:pseudoexp}
\end{figure}

The resonance strength is determined by measuring the unabsorbed photon flux through an absorber layer at height $Z$. The absorption fraction is given by Eq.~\ref{eq:parametrization} and we need to work out the statistic significance of a measurement of the peak's spatial location. Here, let us consider a number of height bins with binwidth $\Delta Z$, and the source provide a flux of  $C_i$ photon arrivals (per unit time) in the $i$th bin, or 
$C_i= \int_{\Delta Z_i} C(Z) dZ$, and we denote the total arrival rate in one bin as $C_{\infty}$ as it should be equal to the $C_i$ far away from the resonance point. For a given \mos source, we choose the binwidth to match its effective height spread under the local gravitational field, namely, $\Delta Z=0.5\cdot g^{-1} \Gamma_{\rm exp}/E_0 $, thus we choose $\Delta Z =10 ~{\rm \mu m}$ for $^{109}$Ag. 
The measured photon flux in the resonance bins will decrease due to resonance absorption, as illustrated by our simulated photon counts in Fig.~\ref{fig:pseudoexp}. As $C_i$ depends on $Z_0$, the location of $Z_0$ is obtained by minimizing the likelihood function with the $\{ C_i^{\rm exp.}\}$ data, 
\be 
\chi^2(Z_0) = \sum_{i}\frac{(C_i(Z_0)-C_i^{\rm exp.})^2}{\Delta_i^2},
\ee
and the spatial resolution of $Z_0$ can be inferred from the likelihood's sensitivity with shifting the $Z_0$ value. A more detailed likelihood would also marginalize over experimental nuisance parameters, which can be calibrated at statistics much higher than the run-time $C_\infty$. For a statistics-dominated estimate, the sensitivity is determined by  $\Delta_i=\sqrt{C_i}$. We empirically obtain the measurement's spatial resolution in $Z_0$ by fitting to simulated data, which translates into the frequency shift sensitivity:
\be 
\frac{\delta f}{f}=\frac{\delta Z_0}{\Delta Z}\cdot\frac{\delta f_{\rm Moss}}{f}\equiv\frac{\xi(\epsilon f_S)}{\sqrt{C_\infty}}\cdot\frac{\Gamma_{\rm exp}}{E_0},
\label{eq:dfoverf}
\ee
for $C_\infty\gg 1$, and the $\xi$ dependence on $\epsilon f_S$ is numerically computed. $f_S$ will depend on the material composition; for metallic silver $f_S=0.05$ and it can be improved by selecting alloys with an higher Debye temperature. For instance, $\rm{AgB_2}$ has $T_{\rm Debye}=408$~K and its $f_S$ is 20\% at 4.2 K~\cite{Ozisik2013TH}.
We performed simulations at different levels of $C_\infty$ with sub-unity values of $f_S$, and the corresponding frequency resolutions are listed in Table~\ref{tbl:sensitivity}. $f_S$ can be increased by  Within the region of interest, $\xi$ can adopt the parametrization:
\be 
\xi(x) = -0.17+0.16x^{-1}+0.014x^{-2}.
\label{eq:empirical_fit}
\ee
In our simulations, we marginalized over two nuisance parameters: the peak width and the peak height. We require sufficient photon counting in the central resonance bins with at least $\sigma=3$ statistical significance: $ C_{\rm res.}\approx C_{\infty}\cdot  f_S \epsilon$, or $ C_{\infty} > (\sigma)^2/(f_S \epsilon)^2$. Clearly, the sensitivity improves over larger $C_\infty$, and the spatial measurement can achieve a fractional resolution of the \mos width with a sufficient source intensity.

\begin{table}[t]
\centering
\begin{tabular}{c|c|c|c|c}
\hline\hline
 Recoil free fraction& \multicolumn{4}{ |c }{\textbf{$C_\infty$}} \\
\hline
   $f_S$ & 50  & 500  & 5000  & 50000  \\
\hline

0.05$^*$  &  -       &  -       &  -       &  1.2e-22  \\
0.10  &  -       &  -       & 1.3e-22  &  3.8e-23  \\
0.20  &  -       & 1.3e-22  & 4.5e-23  &  1.4e-23  \\
0.30  &  -       & 7.9e-23  & 1.9e-23  &  7.0e-24  \\
0.40  &  -       & 4.8e-23  & 1.5e-23  &  4.5e-24  \\
0.50  &  -       & 3.3e-23  & 9.4e-24  &  2.9e-24  \\
0.60  & 7.3e-23  & 2.2e-23  & 7.2e-24  &  2.1e-24  \\
0.70  & 5.0e-23  & 1.5e-23  & 5.0e-24  &  1.5e-24  \\
0.80  & 4.1e-23  & 1.2e-23  & 4.0e-24  & \\
0.90  & 3.7e-23  & 9.5e-24  & 3.1e-24  & \\
\hline\hline
\multicolumn{5}{ l }{$^*$ \text{for metallic silver}} \\



\end{tabular} 
\caption{Simulated frequency shift accuracy ${\delta{f}}/{f}$ achieved with different recoil-free fraction $f_S$ and expected number of gamma-ray counts $C_\infty$ in each 10 $\mu m$ height bin. Scenarios with absorption signal counts less than 3 times of Gaussian fluctuations of expected counts are not listed.
}
\label{tbl:sensitivity}
\end{table}

In case an additional broadening factor applies to $\Gamma_{\rm exp.}\rightarrow B\cdot\Gamma_{\rm exp.}$, both $\Delta Z$ and $C_\infty$ scale linearly with $B$, so that $\delta Z_0$ will scale as $\sqrt{B}$ in Eq.~\ref{eq:dfoverf} due to higher statistics. Thus the overall ${\delta f}/{f}\propto \sqrt{B}$ for a larger $\Gamma_{\rm exp.}$ This is an advantage of resolving the peak-shift: the sensitivity does not degenerate linearly with a wider \mos linewidth. 

\medskip

(2) {\it Time resolution} will determine how frequently the detectors can measure the \mos absorption efficiency, and it sets the maximal gravitational wave signal frequency that our experimental setup can be sensitive to. By Shannon-Nyquist theorem, the minimal sampling frequency needs to be higher than twice of the signal frequency. In order to secure samples close to the maximal signal strength, namely $>90\%$ of $h_0$, the sampling frequency need to be about one order higher. In our estimate, we consider a measuring frequency ten times of that the gravitational wave, or  $10 f_{\rm GW}$. For 10 samples during one period of a sinusoidal waveform, we have around 3 samples of the strain within $90\%-100\%$ of its maximum. 

\begin{table}[t]
    \centering
    \begin{tabular}{l|c|cc|cccc}
    \hline \hline
    &~~$g$~($g_\oplus$)~& ~d (m)~&~ $\Delta Z$ ~& ~$\epsilon f_S$ ~&~ $h_{\rm min}$& $f_{\rm max}$ &$R_s$ (Bq.)\\
    \hline
    ~A~&1 &   1 & 10 ${\rm \mu}$m & 0.04 & $3\times10^{-15}$ & 0.6 KHz & $10^{11}$\\
    ~A'~&1 & 5 & 10 ${\rm \mu}$m & 0.04 & $3\times10^{-17}$ & 13 KHz & $10^{13}$\\
    ~B~&$10^{-4}$ & 10& 1 dm & 0.4 & $3\times 10^{-23}$ & 30 MHz & $10^{14}$\\
    \hline
    ~A$^C$~&1 &  1 & 10 ${\rm \mu}$m & 0.04 & $1\times10^{-21}$ & 3 GHz & $10^{11}$\\
    \hline\hline
    \end{tabular}
    \caption{Sample static \mos measurement configurations that corresponds to a table-top experiment with a Type-III source intensity (A) and a low-$g$ setup with a stronger source (B). A' is scaled-up scenario by increasing the source intensity in A by two orders of magnitude. $h_{\rm min}$ and $f_{\rm max}$ denote the sensitivity to the GW strain and the maximal GW frequency that can be probed. A$^C$ represents the sensitivity with setup A but for a periodic signal with coherence up to $10^6$ periods. The source intensity is given for isotropic sources.}
    \label{tab:benchmark}
\end{table}

\medskip

(3) {\it Counting algorithm.} We can postpone the reconstruction of the signal's $\{\theta,\phi\}$ distribution and sum up the counts from the detectors to obtain a total signal rate. Using the total-count sacrifices the directional information of incident gravitation wave, but in this collective manner it maximizes the sensitivity to the GW magnitude and reduces the requirement on source intensity. As the circle always have two directions perpendicular to the GW's wave-vector, we will consider the counts from detectors located within the region of more than 90\% of maximal signal strength, which has an angular radius $\Delta_{90}$ of more than ten degrees. Thus, we will sum up the counts on the circle within $\pm\Delta_{90}$ sections centering at the maximal $Z$-shift directions. The accurate size of $\Delta_{90}$ will depend on the direction of the incident GW. If the circle happens to locate on a constant $\phi$ plane, $\Delta_{90}$ for $\theta$ is 18.4$^\circ$ by Eq.~\ref{eq:freq_shift}. In case the incident GW is perpendicular to the circle's (horizontal) plane, or $\theta\equiv 90^\circ$, then $\Delta_{90}$ for $\phi$ is $12.9^{\circ}$ and the there are four equal-strength maxima, located at $\phi=0^{\circ}, 90^{\circ}, 180^{\circ}$ and $270^{\circ}$.
With an isotropic \mos source of radioactivity $R_s$, the total photon arrivals rate in one height bin $\Delta Z$ per gravitational wave period is
\be 
N_{90} =  R_s\cdot \frac{2\pi{f}_{t}}{\omega} \cdot  \frac{(2\pi f_{\phi} d)\cdot \Delta{Z}}{4\pi d^2}~,~
\label{eq:N90_arrival}
\ee 
where $d$ is the circle's radius and $f_t\approx 0.3$ is time fraction of the window that samples $>90\%$ of the maximal strain in each signal period. $f_{\phi}$ denotes the fraction of circle within the angular range(s) of $\Delta_{90}$. For incident GWs along the vertical direction, $f_{\phi}=0.288$. Consider Eq.~\ref{eq:empirical_fit}, and identify $N_{90}$ to the binned photon count $C_\infty$ in our pseudo-experiment, we obtain a relation between the frequency shift sensitivity and the source intensity, 
\bea 
R_s &=& \frac{\omega}{2\pi}\frac{C_\infty ~2d}{\Delta Z f_\phi f_t}
    =\frac{2\omega d g \xi^2}{~f_\phi f_t}\left(\frac{\Gamma_{\rm exp}}{E_0}\right)\left(\frac{\delta f}{f}\right)^{-2} \label{eq:source_intensity}\\
    &\approx&10^{14}~{\rm Bq}\cdot\left(\frac{\omega/2\pi}{{\rm MHz}}\right)\left(\frac{d}{1~{\rm m}}\right)\left(\frac{g}{g_\oplus}\right)\nonumber\\
    & &\cdot\left[\frac{\eta(\epsilon f_S)}{12.4}\right]^2\left(\frac{4\times 10^{-21}}{\delta f/f}\right)^2 \nonumber 
\eea
with our detector configurations and $g_\oplus=9.8$ m/s$^2$. Beware that not all the parameters in this formula scale independently.

\medskip
(4) {\it Periodic signals} can benefit from a statistic enhancement by summing up the photon counts during their coherent time scale, rather than only taking into account of $N_{90}$ during one signal period. Recently, narrow-width GW signals have gained strong interest, particularly motivated by coherent collective  behavior of hypothetical low-mass boson fields~\cite{Hui:2016ltb}. A typical dark matter in the galactic halo is expected to have a thermal energy spread around ${\cal O}(10^{-6})$, leading to good coherence over $Q\sim 10^6$ periods. Therefore in case of a coherently repeated signal, an $N_{90}\rightarrow Q\cdot N_{90}$ scaling applies to Eq.~\ref{eq:N90_arrival}, which effectively scales up the source intensity by the same factor $Q$, and significantly boosts the sensitivity to $\Delta f/f$ in its narrow frequency band.

\begin{figure}[t]
    \centering
    \includegraphics[scale=0.61]{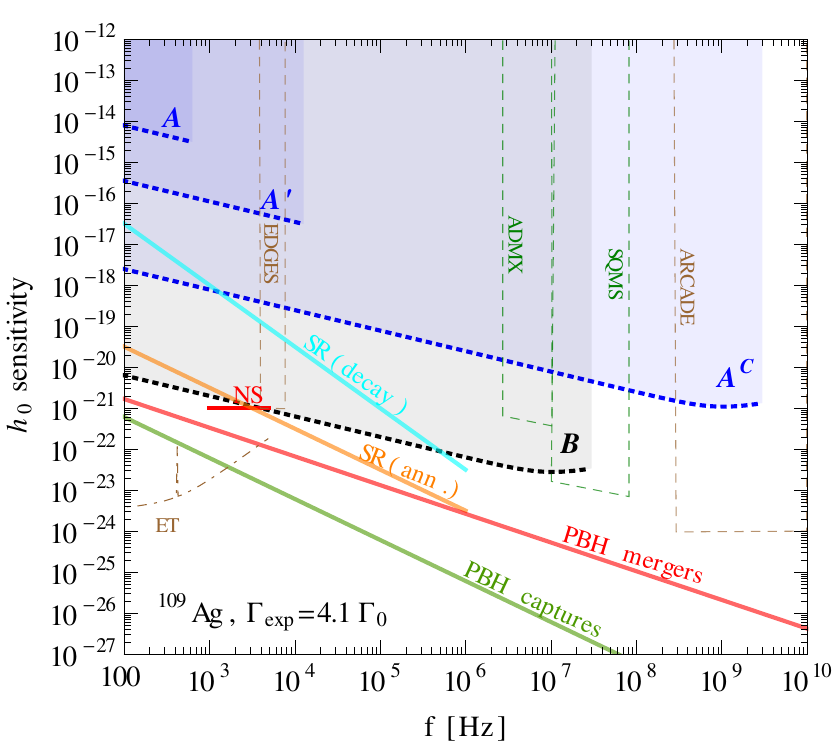}
    \caption{Stationary $^{109}$Ag \mos sensitivity to gravitational wave strain at the benchmark scenarios listed in Table~\ref{tab:benchmark}. The sensitivity curves (dotted) are truncated at an upper frequency $\omega d\sim10$. The theoretical strain predictions from coherent sources, e.g. supernovae (NS), bosonic superradiance (SR) annihilation/decay and primordial black holes (PBH) are shown for comparison. Their strain-frequency predictions are adapted from the recent review on high-frequency GWs~\cite{Aggarwal:2020olq}. The design sensitivity from ET~\cite{LISACosmologyWorkingGroup:2022jok} experiment (dot-dashed) represents the future laser interferometry limits, and EDGES~\cite{Bowman:2018yin} and ARCADE~\cite{Fixsen_2011} sensitivities for those from radio telescopes. The ${\rm A^C}$ curve corresponds to a $Q\sim 10^6$ coherence-enhanced narrow width projection with Scenario A. For comparison, the ADMX~\cite{ADMX:2021nhd} and SQMS~\cite{Herman:2020wao} regions represent their narrow-band sensitivity~\cite{Domcke:2022rgu} based on inverse Gertsenshtein conversion. }
    \label{fig:sensitivity}
\end{figure}

\section{Sensitivity Estimate}
\label{sect:sensitivity}

The GW sensitivity derives from the static measurement's frequency shift sensitivity. Eq.~\ref{eq:source_intensity} shows that for a fixed source intensity, there is a minimal frequency shift that can be experimentally resolved. The corresponding GW strain $h$ sensitivity can be interpreted from Eq.~\ref{eq:freq_shift} below the maximal frequency limit $f_{\rm max}$. Namely, at low frequencies a physical suppression of $(\omega d/2)^{-1}$ applies due to the finite baseline length. The maximal operational frequency $f_{\rm max}$ is determined by the smaller between two frequency cut-offs: (i) reaching the statistics requirement $N_{90}=\sigma^2/(f_S\epsilon)^2$; (ii) reaching $\omega d\sim {\cal O}(10)$, above which the angular pattern of resonance location becomes much more complicated.

In Table~\ref{tab:benchmark}, Scenario A assumes a terrestrial (1$g$) table-top sized experiment with a modest source intensity at $10^{11}$ Bq. For a relatively low-intensity source such as in Scenario A, the statistical $3\sigma$ cut-off (at 0.6 KHz) is much lower than the intrinsic cut-off ${\cal O}(10)\cdot (2\pi d)^{-1}$. Both $f_{\rm \max}$ and the sensitivity $\delta f/f$ are mainly limited by the source intensity. For instance, increasing $R_s$ in Scenario A' to $10^{13}$ Bq. will lift $f_{\rm max}$ to tens of KHz and improve $h_{\rm min}$ to $3\times 10^{-17}$. As $\Delta Z$ does not scale with the baseline length, the angular coverage fraction $f_\phi\Delta Z/2d$ causes a major loss of source efficiency for an isotropic source. Focusing of high-energy photons can be challenging. There are also discussions of X-ray guides~\cite{Vetterling1976MeasurementsOA}\cite{Gonser1981} for efficiency enhancement. We will postpone investigation of non-isotropic sources for later research.

We take note that $\Delta Z$ increases inversely with the local $g$-value, and a low-gravity environment can improve the angular coverage significantly. Scenario B shows a low-gravity setup with a $10^{14}$ Bq source. At $10^{-4}g_{\oplus}$ the height bin increases to around 10 cm, and the strain sensitivity reaches $3\times 10^{-23}$ for a 10 meter radius detector ring. In this setup, the statistics requirement is always satisfied so that the off-cut frequency reaches $f_{\rm max}\sim O(10)\cdot (2\pi d)^{-1}$. The corresponding sensitivity curves are shown as dashed curves in Fig.~\ref{fig:sensitivity}. A larger radius $d$ will let the sensitivity curve to move leftward horizontally, reaching further towards lower GW frequency, at the cost of a stronger source intensity requirement $R_s\propto d$. Given sufficient statistics, the optimal GW frequency for a meter-scale setup is around GHz. A 100 m radius as in Scenario B optimizes for the 10 MHz range, below which the sensitivity decreases linearly with the frequency. As illustrated in Fig.~\ref{fig:sensitivity}, the sensitivity curve covers a wide range of frequencies from KHz to sub-GHz, making the static \mos scheme relevant to potential coherent gravitational wave sources. We would emphasize the A, A' and B sensitivities derive from $N_{90}$ in Eq.~\ref{eq:N90_arrival} that builds on the photon-counting during one signal period. 

In case of a periodic signal with coherence, we also present a narrow-band example by summing the photon numbers over $10^6$ periods, as denoted by Scenario ${\rm A^C}$. The enhanced statistics extend the operational frequency range and allows $f_{\rm max}$ to reach the GHz cut-off with the meter-scale radius. As shown in Fig.~\ref{fig:sensitivity}, a $10^{11}$ source would extend its sensitivity into the GW region that is proposed for electromagnetic  sensors\cite{Pegoraro:1977uv,Caves:1979kq,Reece:1982sc,1988egp..conf..293L,Blair:1995wx}, such as the cavity experiment ADMX~\cite{ADMX:2021nhd} and many others. The interpreted high-frequency limits from some radio telescopes are also shown for comparison. In a similar manner, Scenario B will also extend to deeper strain sensitivity for a coherently periodic signal. The time cost of a narrow-band measurement is much longer, as the integration time at each frequency is also at least $Q$ times of the signal period and the narrow bandwidth slows down the scan rate.

\section{Discussion}
\label{sect:discussion}
To summarize briefly, we have considered a static measurement scheme of \mos resonance that takes advantage of converting the gravitational wave's perturbation into a time-varying vertical displacement of the resonance location. The extreme frequency sensitivity of \mos resonance allows for very promising outlook for gravitational wave detection in the KHz to higher than MHz range with a relatively small-sized ($1\sim 10$  meter) dimension and a radioactive source of reasonable intensity. This provides a promising alternative method of detection in the relatively high-frequency gravitational wave range. With a circular placement of detectors, the static setup has 4$\pi$ coverage of the incoming gravitational wave direction, and it has the potential of resolving the signal direction. The stationary scheme's sensitivity scales improves over a smaller but non-zero local gravity field. A low-gravity environment can significantly boost the experimental reach to gravitational perturbations.

We consider the 88 keV line of $^{109}$Ag as the benchmark \mos isotope. The $^{109}$Ag isotope has a long enough lifetime that offers a practical experimental time scale, and its narrow linewidth guarantees a reasonable absorber/detector size under the terrestrial gravitational field. Generally speaking, the narrower line-width the better sensitivity. At a fixed source intensity, the stationary scheme's spatial sensitivity scales only as the square-root of the effective \mos linewidth. The choice of isotope needs to balance between the sensitivity, the mother isotope's lifetime and the vertical shift length under the local gravity field. In low-$g$ environments, isomers with even sharper linewidth, such as $^{103}$Rh and $^{189}$Os, could be interesting options if their short lifetime issue can be solved.

In perspective, one can also think of increasing the photon counting statistics from the \mos source and optical enhancement. From the source side, it was proposed to amplify the source intensity through stimulated emission of the ensemble of nuclei in a host dielectric crystal using laser irradiation~\cite{Tkalya:2010df}. 
Recently, femotosecond pumping of nuclear isometric states at 41.6 keV and 562.5 keV $\gamma $-ray of $^{83}$Kr has been demonstrated using 30 fs laser pulses at 120 TW~\cite{PhysRevLett.128.052501}. Based on these developments, it is promising to consider a nuclear laser at the resonance energy of $^{109}$Ag. In addition, focusing using refractive lens~\cite{RN5055} and multi-layer Laue lens~\cite{RN2796} have been recently demonstrated at synchrotron radiation facilities. Any focusing of the 88 keV $\gamma $-ray of $^{109}$Ag emitted from the source will increase the beam density at the detectors.

\bigskip
{\bf Acknowledgements.}\\
~~Authors thank Kai Liu and Wanquan Shi for helpful communications. Y.G. is supported in part by the National Natural Science Foundation of China (no. 12150010 and no. 12275278). H.Z. is supported by the Ministry of Science and Technology of China (no. 2022YFA1602100) and Natural Science Foundation of China (no. 12061141003). W.X. is supported by the High Energy Photon Source (HEPS), a major national science and technology infrastructure in China, and by the National Natural Science Foundation of China (no. 12075273).


\bibliography{refs}

\end{document}